\documentclass[aps,prb,showpacs,twocolumn]{revtex4}
\usepackage{amssymb}
\usepackage{graphicx}

\begin{document}

\title{Einstein's mirror revisited\footnote{E-print arXiv:physics/0701222}}

\author{Aleksandar Gjurchinovski}

\email{agjurcin@iunona.pmf.ukim.edu.mk}

\author{Aleksandar Skeparovski}

\email{skepalek@iunona.pmf.ukim.edu.mk}

\affiliation{Department of Physics, Faculty of Natural Sciences 
and Mathematics, Sts.\ Cyril and Methodius University,
P.\ O.\ Box 162, 1000 Skopje, Macedonia}

\date{March 07, 2007}

\begin{abstract}
We describe a simple geometrical derivation of the formula for reflection of
light from a uniformly moving plane mirror directly from the postulates of 
special relativity.
\end{abstract}

\pacs{03.30.+p}

\maketitle

\section{Introduction}

Reflection of light from a plane mirror in uniform rectilinear motion is a 
century-old problem, intimately related to the foundations of special relativity.\cite{michelson,shankland,abraham,einstein,majorana,kennard,kennedy1,kennedy2,soni1,
soni2,schumacher} The reflection formula is usually obtained by referring to the 
Lorentz transformation to switch from the mirror's rest frame to the frame where the 
mirror moves at a constant velocity.\cite{einstein,bolotovskii,stephani,
pauli,synge,yeh} The reflection formula also follows from the constancy of the 
speed of light, by using Huygens' construction\cite{gjurchinovski1} or 
Fermat's principle of least time\cite{gjurchinovski2}. 

In this paper, we introduce a derivation of the reflection formula by comparing 
the geometry of the problem in the frame where the mirror is moving to the one in 
the frame where the mirror is stationary. We analyze three distinct cases of the 
mirror's motion, when the mirror is moving: a) parallel to its surface; b) perpendicular to 
its surface; and c) in an arbitrary direction. The derivation requires nothing but a simple 
plane geometry and an elementary understanding of the postulates of special relativity, 
and does not use Lorentz transformation or any additional tools of classical wave optics.

\section{Reflection from a plane mirror moving uniformly parallel to its surface}

A plane mirror of length $2\ell=\overline{LR}$ is placed horizontally in its 
rest frame (see Fig. 1). 
\begin{figure}
\includegraphics[width=.8\columnwidth,angle=0,clip]{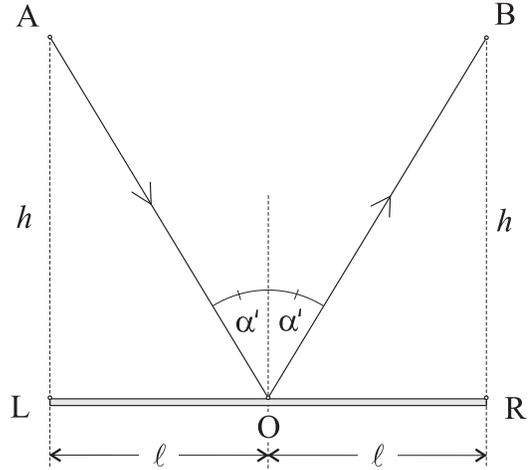}
\caption{Reflection of the photon from a horizontal mirror with respect to the 
mirror's rest frame.}
\end{figure}
A photon is emitted from a source $A$ located at a vertical distance $h$ from the left 
edge $L$ of the mirror. The photon bounces off the midpoint $O$ from the mirror, and 
it is absorbed by the detector $B$ located at the same vertical distance $h$ from 
the right edge $R$ of the mirror. Evidently, the angle of incidence and the angle 
of reflection of the photon are equal.

We now transfer to the frame in which the mirror, the source and the detector are all 
moving at a constant speed $v$ to the right (see Fig. 2). 
\begin{figure}
\includegraphics[width=.8\columnwidth,angle=0,clip]{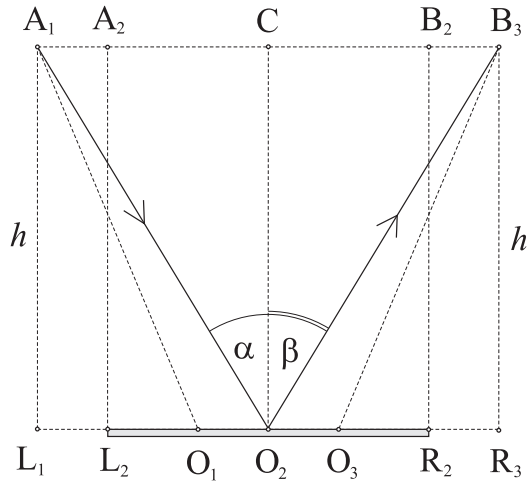}
\caption{The situation in Fig. 1 with respect to an observer moving at a constant 
speed $v$ to the left.}
\end{figure}
Since the point of reflection in the mirror's rest frame divides its length into two 
equal halves, it must also divide the mirror in two equal halves with respect to any other 
inertial frame.\cite{mermin} By making this claim, we implicitly invoked the principle of 
relativity, according to which one cannot distinguish one inertial frame from another. 
Also, at each instant of time, the moving source and the moving detector are located at 
a vertical distance $h$ above the left and the right edges of the moving mirror, 
respectively, as they were in the frame where the mirror was stationary. Here, we again 
appealed to the principle of relativity which directly implies the invariance of the 
lengths measured perpendicularly to the direction of relative motion between the frames.\cite{young}  

In Fig. 2 we see that the photon was emitted from the moving source at the time when the source 
was located at $A_1$. At this instant, the left edge of the mirror was located at point $L_1$ 
at a vertical distance $h$ below $A_1$, and $O_1$ is the location of the midpoint of the mirror. 
The photon bounces off the midpoint $O_2$ a time $t_1$ from its emission at $A_1$. 
During this time $t_1$, the photon has traveled the path $\overline{A_1O_2}=ct_1$, 
and the mirror, the source and the detector have moved the distance $vt_1$ to the 
right. We have used Einstein's second postulate that the photon will move at a speed of 
light $c$ with respect to every inertial frame of reference.

During the time $t_2$ from the reflection, the photon has 
traversed the path $\overline{O_2B_3}=ct_2$ being absorbed by the detector at point 
$B_3$. Accordingly, the mirror, the source and the detector have moved the additional 
distance $vt_2$ to the right. At the time of the absorption, the right 
edge of the mirror was located at point $R_3$ at a distance $h$ vertically below 
$B_3$, and $O_3$ is the location of the midpoint of the mirror. If we denote by $\alpha$ the 
angle of incidence, and by $\beta$ the angle of reflection of the photon, and 
note the triangle similarities in Fig. 2, we may write:
\begin{eqnarray}
\alpha&=&\measuredangle{A_1O_2C}=\measuredangle{L_1A_1O_2} \nonumber \\
&=&\measuredangle{L_1A_1O_1}+\measuredangle{O_1A_1O_2},
\label{eq:2.1}\\
\beta&=&\measuredangle{CO_2B_3}=\measuredangle{O_2B_3R_3}  \nonumber \\
&=&\measuredangle{O_2B_3O_3}+\measuredangle{O_3B_3R_3}.
\label{eq:2.2}
\end{eqnarray}
Obviously, $\triangle{A_1L_1O_1}=\triangle{B_3R_3O_3}$, and thus:
\begin{equation}
\measuredangle{O_3B_3R_3}=\measuredangle{L_1A_1O_1}.
\label{eq:2.3}
\end{equation}
Applying the sine theorem to $\triangle{O_1A_1O_2}$, we obtain:
\begin{equation}
\sin{\measuredangle{O_1A_1O_2}}={\overline{O_1O_2}\over \overline{A_1O_2}}
\sin{\measuredangle{A_1O_1O_2}}.
\label{eq:2.4}
\end{equation}
Since $\overline{O_1O_2}=vt_1$ and $\overline{A_1O_2}=ct_1$, we get:
\begin{equation}
\sin{\measuredangle{O_1A_1O_2}}={v\over c}
\sin{\measuredangle{A_1O_1O_2}}.
\label{eq:2.5}
\end{equation}
Analogously, for $\triangle{O_2B_3O_3}$ we obtain:
\begin{equation}
\sin{\measuredangle{O_2B_3O_3}}={v\over c}
\sin{\measuredangle{O_2O_3B_3}},
\label{eq:2.6}
\end{equation}
where we have taken into account that $\overline{O_2O_3}=vt_2$ and $\overline{O_2B_3}=ct_2$.
But $\measuredangle{A_1O_1O_2}=\measuredangle{O_2O_3B_3}$, which we use into Eqs. (\ref{eq:2.5})
and (\ref{eq:2.6}) to get:
\begin{equation}
\measuredangle{O_2B_3O_3}=\measuredangle{O_1A_1O_2}.
\label{eq:2.7}
\end{equation}
From Eqs. (\ref{eq:2.1}), (\ref{eq:2.2}), (\ref{eq:2.3}) 
and (\ref{eq:2.7}) we deduce $\alpha=\beta$, that is, the incident angle of the photon equals
the reflected angle. Thus, the motion of the mirror in its plane will not affect the reflection of 
the photon, and the photon will be reflected in accordance with the usual law of reflection as if
the mirror were stationary.

\section{Reflection from a plane mirror moving uniformly perpendicular to its surface}

In this section we consider the reflection of the photon when the velocity of the mirror is normal 
to its surface. The schematic of the reflection with respect to the mirror's rest frame is shown
in Fig. 3. 
\begin{figure}
\includegraphics[width=.8\columnwidth,angle=0,clip]{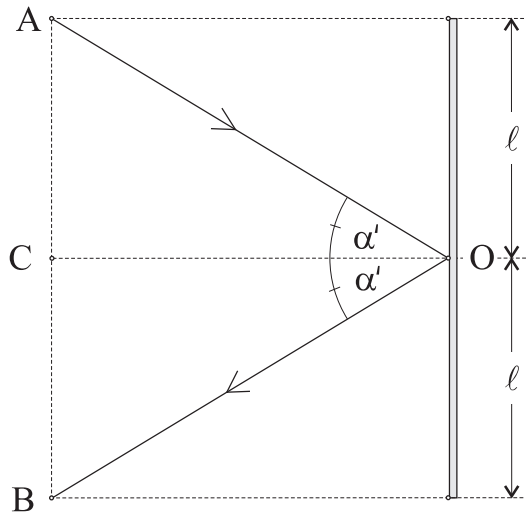}
\caption{Reflection at a vertical mirror observed from a frame where the mirror is stationary.}
\end{figure}
The photon is emitted from a fixed source at $A$, bounces off the mirror at its midpoint $O$, 
and eventually hits the fixed detector at $B$. The source $A$ and the detector $B$ are located at a 
vertical line $ACB$ parallel to the mirror's surface, being equally distant, but at the opposite sides, 
from the line $OC$ normal to the mirror's surface ($\overline{AC}=\overline{BC}=\ell$). The source 
$A$ is horizontally aligned with the upper edge of the mirror, and the detector $B$ with its 
lower edge. A simple trig reveals that the photon will be reflected off the mirror at the same angle 
at which it was incident.

The situation with respect to the frame where the mirror, the source and the detector are moving at 
a constant speed $v$ to the right is shown in Fig. 4. We will apply the same arguments as in the previous 
section to derive the formula for reflection of the photon in this case. 
\begin{figure}
\includegraphics[width=.8\columnwidth,angle=0,clip]{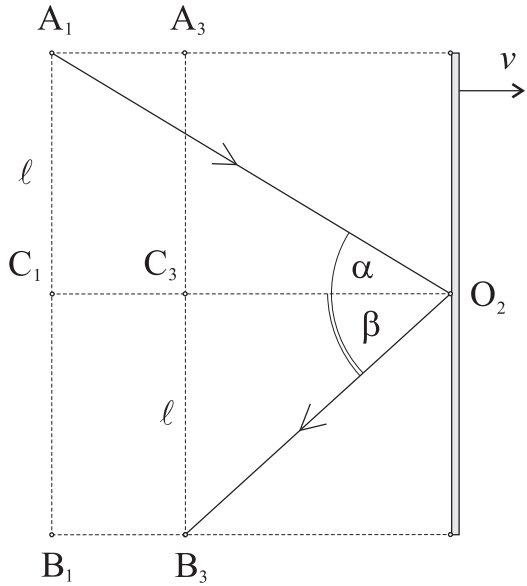}
\caption{The situation in Fig. 3 observed from a reference frame moving at a constant 
speed $v$ to the left.}
\end{figure}

After being emitted from the moving source at time when the source was located at $A_1$, the photon 
is reflected from the moving mirror off its midpoint $O_2$ and then absorbed by the moving detector 
at the point $B_3$. If $t_1$ is the photon's transit time from $A_1$ to $O_2$, and $t_2$ the time from 
$O_2$ to $B_3$, we have $\overline{A_1O_2}=ct_1$ and $\overline{O_2B_3}=ct_2$, where we have taken into 
account the constant speed of light postulate. At time when the photon reaches the detector at $B_3$, 
the detector has moved the distance $\overline{B_1B_3}=v(t_1+t_2)$ to the right from its position 
$B_1$ when the photon was emitted from $A_1$. From $\triangle{A_1C_1O_2}$ and $\triangle{B_3C_3O_2}$ 
in Fig. 4, we have:
\begin{eqnarray}
\cos\alpha&=&{b\over ct_1}, \label{eq:3.1} \\
\cos\beta&=&{b-v(t_1+t_2)\over ct_2}, \label{eq:3.2}
\end{eqnarray}    
where $\alpha$ and $\beta$ are the incident angle and the reflected angle of the photon, respectively,
and $b=\overline{C_1O_2}$. We eliminate $b$ from Eqs. (\ref{eq:3.1}) and (\ref{eq:3.2}) to get:
\begin{equation}
\cos\beta=\left(\cos\alpha-{v\over c}\right){t_1\over t_2}-{v\over c}.
\label{eq:3.3}
\end{equation}
We apply the Pythagoras theorem to $\triangle{A_1C_1O_2}$ and $\triangle{B_3C_3O_2}$ to obtain:
\begin{eqnarray}
(ct_1)^2&=&\ell^2+b^2, \label{eq:3.4} \\
(ct_2)^2&=&\ell^2+\left[b-v(t_1+t_2)\right]^2 \label{eq:3.5},
\end{eqnarray}
where $\ell=\overline{A_1C_1}=\overline{B_3C_3}$.
By subtracting Eq. (\ref{eq:3.5}) from Eq. (\ref{eq:3.4}) to eliminate $\ell$, we obtain:
\begin{equation}
(ct_1)^2-(ct_2)^2=b^2-\left[b-v(t_1+t_2)\right]^2.
\label{eq:3.6}
\end{equation}
The last equation can be factored into the form:
\begin{equation}
c^2(t_1-t_2)(t_1+t_2)=v(t_1+t_2)\left[2b-v(t_1+t_2)\right],
\label{eq:3.7}
\end{equation}
which upon division by $(t_1+t_2)$ and using Eq. (\ref{eq:3.1}) can be recasted into:
\begin{equation}
{t_1\over t_2}={1-v^2/c^2\over 1-2(v/c)\cos\alpha+v^2/c^2}.
\label{eq:3.8}
\end{equation}
We substitute the last expression for $t_1/t_2$ into Eq. (\ref{eq:3.3}) to obtain:
\begin{equation}
\cos\beta={-2v/c+(1+v^2/c^2)\cos\alpha \over 1-2(v/c)\cos\alpha+v^2/c^2}.
\label{eq:3.9}
\end{equation}
The result gives the angle of reflection $\beta$ of the photon in terms of the incident 
angle $\alpha$ and the speed $v$ of the mirror when the mirror is moving perpendicularly to 
its surface. It exactly matches the reflection formula obtained otherwise.\cite{einstein,
bolotovskii,yeh,gjurchinovski1,gjurchinovski2}
\begin{figure}
\includegraphics[width=1\columnwidth,angle=0,clip]{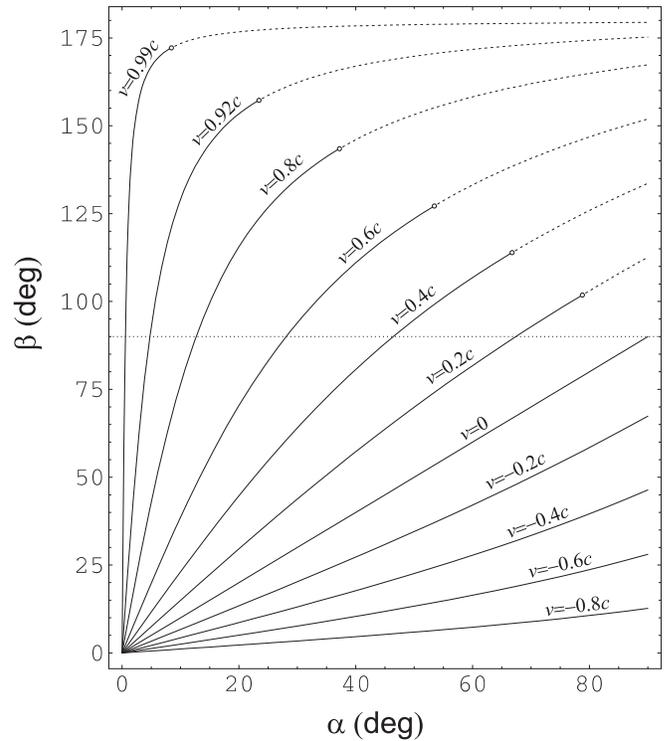}
\caption{The reflection angle $\beta$ versus the incident angle $\alpha$ of the photon
for different values of the speed $v$ of the mirror moving perpendicularly to its surface.
The portions of the curves corresponding to $v>0$ and $\alpha\geq\arccos(v/c)$ are drawn with dashed 
lines. At incident angles corresponding to the dashed curves, the reflection does not occur.}
\end{figure}

The behavior of the reflected angle $\beta$ as a function of $\alpha$ for different values 
of the speed $v$ of the mirror is given in Fig. 5. Note that the negative 
values of $v$ correspond to the motion of the mirror in the opposite direction to 
the one given in Fig. 4. From Fig. 5 it is evident that the reflected photon no longer
obeys the usual law of reflection, except when $v=0$ which is the case when the mirror is
stationary. It is also evident that $\beta<\alpha$ when $v<0$, and $\beta>\alpha$ when $v>0$.

An interesting property of the formula (\ref{eq:3.9}) is observed for the case $v>0$. Namely,
for a given positive value for the speed $v$ of the mirror, there exist an interval of values 
for the incident angle $\alpha$ of the photon $(\alpha_c<\alpha<\alpha_{\text{max}})$, for which the photon 
is reflected at angles larger than $90^\circ$. Hence, the reflected photon, instead of moving away from the 
mirror, moves in the same general direction as the mirror. The effect is known as "the forward 
reflection of light".\cite{jaffe}
The critical angle $\alpha_c$ at which the forward reflection begins can be calculated from 
the formula:
\begin{equation}
\cos\alpha_c={2v/c\over 1+v^2/c^2},
\label{eq:3.10}
\end{equation}
which follows from Eq. (\ref{eq:3.9}) by applying the condition $\beta=90^\circ$. The forward reflection
stops when the incident angle attains the value $\alpha_{\text{max}}$, satisfying the formula: 
\begin{equation}
\cos\alpha_{\text{max}}=v/c.
\label{eq:3.11}
\end{equation}
When $\alpha=\alpha_{\text{max}}$, the photon's velocity component along the motion of 
the mirror ($c\cos\alpha$) will match the speed $v$ of the mirror. Hence, the photon will never reach 
the mirror's surface, and the reflection will never occur. The conclusion remains the same 
for incident angles larger than $\alpha_{\text{max}}$. In this case, the mirror's velocity exceeds 
the velocity component of the photon in the mirror's moving direction. Translating to the frame
in which the mirror is at rest, and thus taking into account the aberration
phenomenon, the situation corresponds to a photon ``incident'' on the mirror at angles
larger than $90^\circ$. The portions of the curves in 
Fig. 5 corresponding to $\alpha\geq \alpha_{\text{max}}$ are drawn with dashed lines. 

\section{Reflection from a plane mirror moving in an arbitrary direction at a constant velocity}

We may generalize the above approach to the case when the reflection occurs off the mirror 
moving at a speed $v$ directed at an angle $\varphi$ from its surface normal.
Since the motion of the mirror in its plane does not affect the reflection (see Sec. II), the 
reflection formula in this case follows from Eq. (\ref{eq:3.9}) in Sec. III  if we simply 
replace $v$ by $v\cos\varphi$, which is the velocity component of the mirror that is normal to its surface:
\begin{equation}
\cos\beta={-2(v/c)\cos\varphi+[1+(v^2/c^2)\cos^2\varphi]\cos\alpha \over 1
-2(v/c)\cos\alpha\cos\varphi+(v^2/c^2)\cos^2\varphi}.
\label{eq:4.1}
\end{equation}
When $\varphi=0$, the motion of the mirror is perpendicular to its surface, and Eq. (\ref{eq:4.1}) 
reduces to  Eq. (\ref{eq:3.9}). Also, when $\varphi=90^\circ$, the mirror moves parallel to 
its surface, and the reflection formula simplifies to $\alpha$=$\beta$. 

The detailed analysis of Eq. (\ref{eq:4.1}) for different values of $\varphi$ is left to the student 
as an exercise.

\begin{acknowledgments}
We thank Iacopo Carusotto (University of Trento) for a fruitful discussion on a similar problem 
that substantially improved the conclusions associated with Fig. 5, and Vase Jovanov 
(TU Munich) for helping us with the references.
\end{acknowledgments}

\end{document}